\documentclass[preprintnumbers,amsmath,amssymb,twocolumn,aps,pre]{revtex4-1}

\usepackage{amssymb}
\usepackage{color}
\usepackage{amsmath}
\usepackage{mathrsfs}
\usepackage{graphicx}
\usepackage{subfig}
\usepackage{epsfig}
\begin{document}

% Title Page
\title{Effects of dynamic-demand-control appliances on the power grid 
frequency}
\author{E.B. Tchawou~Tchuisseu$^1$, D. Gomila$^1$,  D. 
Brunner$^{1,2}$ and P. Colet$^1$}
\affiliation{$^1$Instituto de F\'{\i}sica Interdisciplinar y Sistemas Complejos, IFISC (CSIC-UIB), 
Campus Universitat Illes Balears, E-07122 Palma de Mallorca \\
$^2$D\'epartement Optique Institut FEMTO-ST UMR 6174 - Univ. Bourgogne Franche-Comt\'e CNRS, 25030
Besancon cedex, France}
\date{\today}

\begin{abstract}
Power grid frequency control is a demanding task requiring expensive idle power plants to adapt the supply to the fluctuating demand. An alternative approach is controlling the demand side in such a way that certain appliances modify their operation to adapt to the power availability. This is specially important to achieve a high penetration of renewable energy sources.  
A number of methods to manage the demand side have been proposed. In this work we focus on dynamic demand control (DDC), where smart appliances can delay their switchings depending on the frequency of the system. We introduce a simple model to study the effects of DDC on the frequency of the power grid. The model includes the power plant equations, a stochastic model for the demand that reproduces, adjusting a single parameter, the statistical properties of frequency fluctuations measured experimentally, and a generic DDC protocol. We find that DDC can reduce small and medium size fluctuations but it can also increase the probability of observing large frequency peaks due to the necessity of recovering pending task. We also conclude that a deployment of DDC around 30-40\% already allows a significant reduction of the fluctuations while keeping the number of pending tasks low.
\end{abstract}

\maketitle

\section{Introduction}
During the two past decades several studies have been carried out on how to 
control, stabilize and improve the efficiency of electricity power grid 
\cite{Lasseter1,Lasseter2,Nehrir,Short07,Brooks,Lu,Takagi,Molderink,Kosek}. 
%The electric system is divided in four main parts: generation, transmission, distribution, and consumption. Generation and consumption have taken, in general, the most part of the attention. 
Electric fluctuations have two main contributions. On the side 
of demand, households and commercial users attached to a grid do not have ideal constant or periodic needs. A significant amount of devices are switched on and off at an arbitrary time over the day. 
On the side of generation, although the majority of the production is provided by nuclear and fossil power plants, renewable energy sources account already a significant fraction of the production (over 20\% in Spain, for instance) and there is an overall growing interest in their use. Renewables are 
also constantly subjected to fluctuations at different time scales, for 
instance, turbulence of the wind on windmills, clouds passing over 
photovoltaic panels, etc. The fluctuations on the demand side combined with the 
fluctuations of the production due to renewable sources, constantly unbalance 
production and demand, creating fluctuations of the frequency of the system.
We note that in some instances a group of households may decide to switch on a device at the 
same time, due to for example a TV program break. Indeed the largest 
'TV Pickup' recorded in UK was about $2.8$ GW at the end of the 1990 
World Cup semi-final between England and West Germany \cite{Trevor}.
Here we are concerned on random fluctuations. Large fluctuations triggered
by external factors will not be considered in this work.

In general, electricity generated by a power plant and consumed in households and commercial users 
is characterized by three main properties: voltage, current and 
frequency. Despite some proposals \cite{Takagi,Shively}, there is practically no electricity storage capacity in the power grid, therefore the power generated at 
any time has to match exactly the power consumed by all the loads attached to the grid \cite{Short07}.
Thus current is expected to vary according consumption. However, voltage and frequency must be kept within acceptable ranges by the system operator as established by the legislation of each country. For instance in most part of Europe, the grid is designed to run at $50 \pm 0.5$ Hz and tension variations below $7\%$ \cite{Parliament,National}. Eventual mismatches between supply and demand modify the frequency of the grid. In fact, if at any time power demand exceeds supply, the frequency falls (associated to turbine rotation slow down). Frequency is then a good proxy to monitor the stability of the power grid. 

Based on the mismatch between supply and demand, due to the fact that 
turbine governors can not follow fluctuations in demand fast enough,
several approaches to control the demand side the grid fluctuations have been proposed. One 
is Dynamic Demand Control (DDC), which provides frequency 
regulation by controlling the demand side of the grid, reducing the load when 
the grid is under stress and increasing it when there is a surplus generation \cite{Short07,Moghadam}. 
While stabilizing the power grid, DDC aims at making it more resilient against power outages, as well as  
saving spinning reserve availability. DDC can be implemented as an 
external control or integrated within the electric appliances \cite{Short07,Hild}.
DDC algorithms can work in rechargeable devices such a laptops or mobile 
phones, or with appliances for thermal applications like  air conditioners, 
refrigerators, electric boilers, etc. These devices sometimes can delay or 
advance their operation minutes or even hours without disrupting the user 
comfort.

Besides DDC, other works proposing demand-side control are the following: in 1979 
Schweppe et al in \cite{Schweppe} proposed the concept of ``Frequency 
Adaptive Power energy Re-scheduler'' (FAPER).
They proposed the idea of individual load control through responding 
to frequency. Econnect Ltd developed a 'Distributed Intelligent Load 
Controller' using additional loads to handle excess generation and load 
shedding to handle shortages based on frequency sensing using fuzzy logic 
\cite{Pandiaraj}. The UK firm ResposiveLoad Ltd has
developed a frequency-dependent load controller, which uses various frequency 
limits to affect the probability to switching \cite{Hirst}. 
Recently a dynamic demand response where the 
price of electricity is directly linked to the frequency has been considered 
\cite{Timme}.

One of the issues of using DDC applied to 
domestic appliances is their synchronization. The algorithm continuously monitors the electrical 
frequency shutting off the appliance load when the electrical frequency drops below a lower threshold and 
remaining in the off state until the frequency goes above an upper threshold \cite{Parliament}. 
If all DDC-controlled appliances operate with the same threshold, the simultaneous responses 
can lead to oscillatory instabilities in the frequency  \cite{Short07,Parliament}. To address this issue,
randomization of the action of each appliance has been suggested 
\cite{Short07,Parliament,Mohsenian,Saadat}. 

In this work we study a generic algorithm of DDC to study the effects of 
this technique on the power grid fluctuations. To do so, we first introduce a simple
stochastic model for the demand such that when coupled to a standard 
power plant model it is capable of reproducing the statistical properties of the 
frequency fluctuations measured experimentally. We then introduce DDC 
on the appliances and analyze its effect on the frequency fluctuations as well as its 
efficiency in stabilizing the electrical power grid. We find 
that, DDC can significantly reduce small and medium size fluctuations, however, the 
recovery of pending tasks may increase the probability of large frequency peaks. 
Finally we consider that only a fraction of the load of the system is 
controlled by smart appliances and study how this fraction affects the 
overall performance of the power grid. 

The paper is organized as follows. In Sect.~\ref{sec_powerplant} we describe 
the standard power plant model to be considered in this paper. In Sect.~\ref{sec_demand} we introduce 
the stochastic model for demand which is calibrated in Sect.~\ref{results_demand} on the 
basis of experimental measurements of the frequency fluctuations. 
In Sect.~\ref{sec_ddc} we introduce the model for dynamic demand control. In 
Sect.~\ref{results_DDC} we study the effects of DDC on frequency fluctuations and 
determine the most suitable operation parameter values. In Sect.~\ref{sec_fraction} 
we analyze the how the effectiveness of DDC depends on the number of smart 
devices in the grid. Finally, in Sect.~\ref{conclusions}, we give some 
concluding remarks. 

\section{Power Plant Model}
\label{sec_powerplant}

A conventional power plant is, roughly speaking, constituted by a 
generator and a governor. The generator is the responsible to produce electricity 
out of a fuel or a renewable energy source, and the governor is the 
specific control method used to match the power of the generator to the demand. 
The generator is typically composed by a 
mechanical part, often a turbine fixed on a ferromagnetic rotor which 
rotates between the stator winding, and an electrical part formed by coils 
wound around the stator. Both parts 
are coupled magnetically. By applying the Newton law on the turbine, the 
well-known swing equation describing the dynamics of the generator can be 
derived \cite{Saadat}: 
\begin{equation}  
 \frac{d\omega}{dt}=\frac{\omega}{2HP_{\rm G}}(P_{\rm m}-P_{\rm e}),
 \label{eq1} 
\end{equation} 
where $P_{\rm G}$ and $H$ are the nominal capacity and the inertia constant of the generator
$P_{\rm m}$ is the mechanical power generated by the turbine (or other means),
and $P_{\rm e}$ is the total power of the electric current passing 
through the coils around the stator. The total electric load $P_{\rm e}$ can be divided in two parts:  a 
non frequency-sensitive load and frequency-sensitive load, such that
\begin{equation}  
 P_{\rm e}(\omega,t)=\left(1+ D \frac{\omega-\omega_{\rm R}}{\omega_{\rm R}}\right) P(t),
 \label{pfs} 
 \end{equation}
where $\omega_{\rm R}$ is the grid reference frequency, $D$ is the fraction of the load 
which is frequency sensitive, such as electrical motors, and $P(t)$ is the 
total power at $\omega=\omega_{\rm R}$.

In Eq.~(\ref{eq1}) if the electrical power exceeds the input mechanical power, for instance as a consequence of a sudden load increase, the frequency of the system decreases. Conversely, if the input mechanical power exceeds the load, the 
frequency increases. The governor is responsible to restore the 
frequency to its reference value, and it does so in two steps.
 Under a supply-load unbalance, the primary regulation acts within tens 
of seconds increasing (or decreasing) the mechanical power to halt the decline 
(or rise) in frequency. The secondary 
regulation, which acts within tens of minutes, incorporates 
spinning reserve to the generation in order to restore the frequency to its 
reference value $\omega_{\rm R}$ \cite{Saadat,Short07}. Secondary regulation is conditioned to the availability of sufficient spinning reserved. 
Primary and secondary regulations can be respectively modeled 
by the following equations:
\begin{eqnarray}
\frac{dP_{\rm m}}{dt}&= 
\frac{1}{\tau_{\rm g}}[P_{\rm s}-P_{\rm m}-\frac{P_{\rm G}}{R\omega_{\rm R}}(\omega-\omega_{\rm R})]
\label{eq2}  
\\
\frac{dP_{\rm s}}{dt}&= -\frac{K}
{\omega_{\rm R}}(\omega-\omega_{\rm R}). 
\label{eq3}  
\end{eqnarray}
Here $R$ is the governor speed regulation, $P_{\rm s}$ is the spinning reserve power used at a given time, $K$ is the gain of the secondary controller and $\tau_{\rm g}$ is the time constant of the turbine. 

\begin{figure}
\includegraphics[width=0.5\textwidth]{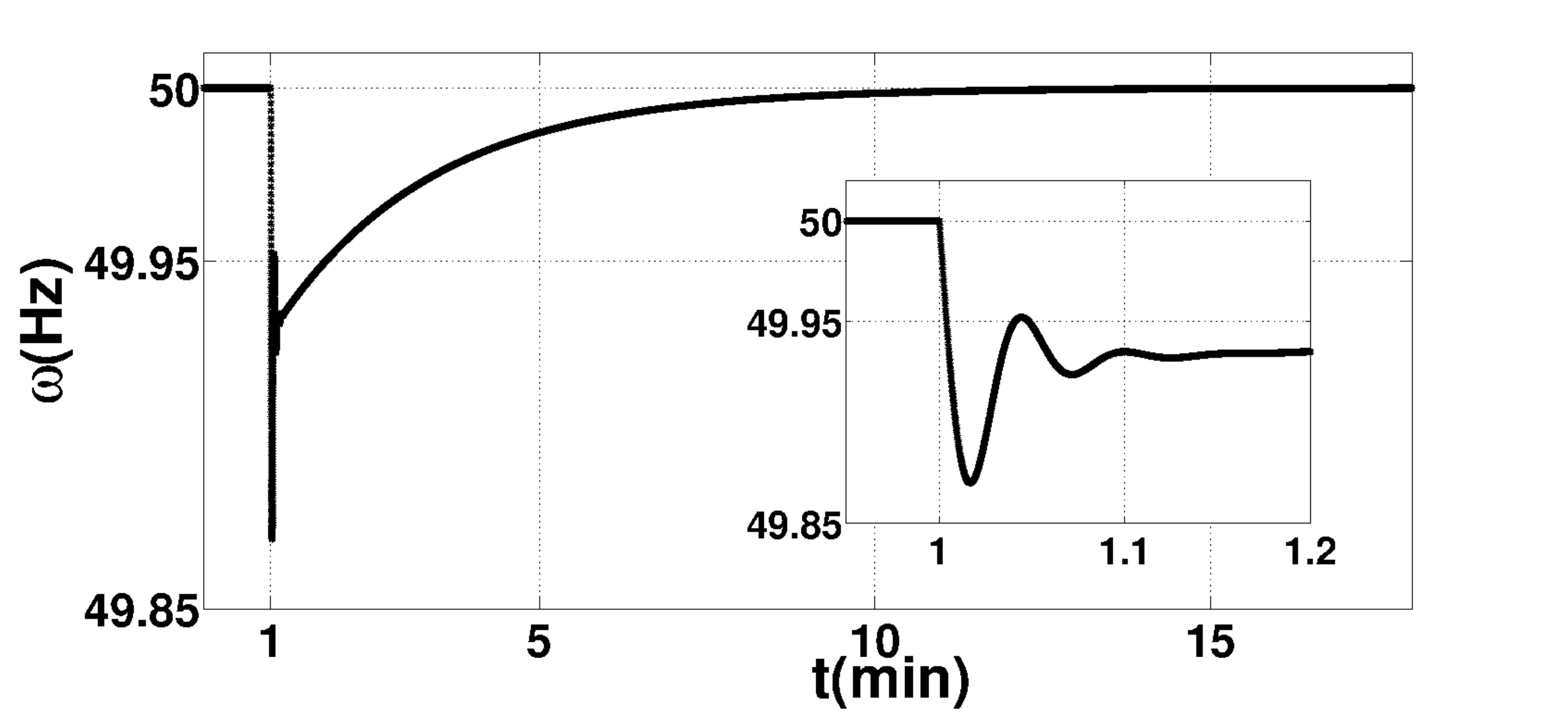}
\caption{Frequency response after a sudden increase of the demand. Initially 
the system is operating at a constant load of $36000$ MW. At $t=1$ min a 
sudden increase of load of $1320 MW$ is applied. Other parameters are 
$\omega_{\rm R}=50$ Hz, $P_{\rm G} = 37320$ MW, $\tau_{\rm g}=0.78$ s, $H=4s $, $D=0.026$, $R=0.04$ and $K=6600$ MW/s.}
\label{fig1}
\end{figure}

Fig.~\ref{fig1} illustrates the response of the system after a sudden load increase that takes place at time $t=1$ minute.
Immediately after the load increase the frequency decreases. The first response of the power plant due to the governor, 
Eq.~(\ref{eq2}), acts to stop the frequency decline, and within tens of second the frequency is momentarily stabilized at a 
value below $\omega_{\rm R}$, ($\omega=49.93$ Hz). After 
stabilizing the decline, secondary regulation, Eq.~(\ref{eq3}) returns the frequency to its reference value $\omega_{\rm R}=50$  
Hz within about 12 min, which corresponds to a realistic power plants response time \cite{Short07}.

\section{Stochastic demand model}
\label{sec_demand}

We propose a very simple model able to reproduce the main statistical 
properties of real demand fluctuations by fitting a single parameter. This will 
allow us to study general effects of applying DDC to the power grid in the next 
section. To model a load consisting of generic domestic appliances such as 
refrigerators, freezers, air conditioners, electric heaters, dishwashers, 
chargeable portable devices, etc, that can switch on and off at any time, we 
consider $N$ devices, or bunches of aggregated devices. The total load demand will be
\begin{equation}  
 P(t)=\sum_{j=1}^{N}P_{j}(t), 
 \label{prandom} 
 \end{equation}
where $P_{j}(t)$ is the load of device (or group of devices) $j$ at time $t$.

For the sake of simplicity in this work we consider that $P_{j}(t)$ 
can only take the values $0$ (off) or $P_{0}$ (on) \cite{footnote1}. 
We consider that appliances in the off state switch on with a rate $p$, while running devices switch off with rate $q$.  
This creates a fluctuating demand with statistical 
properties that depend on the parameters $p$, and $q$. Throughout this work we 
will consider that the rates are constant and identical, $p=q$, such that the average power remains constant. Time varying 
rates $p(t)$, $q(t)$ following daily demand patterns will be considered elsewhere.    

\begin{figure} 
\centering
\includegraphics[width=0.5\textwidth]{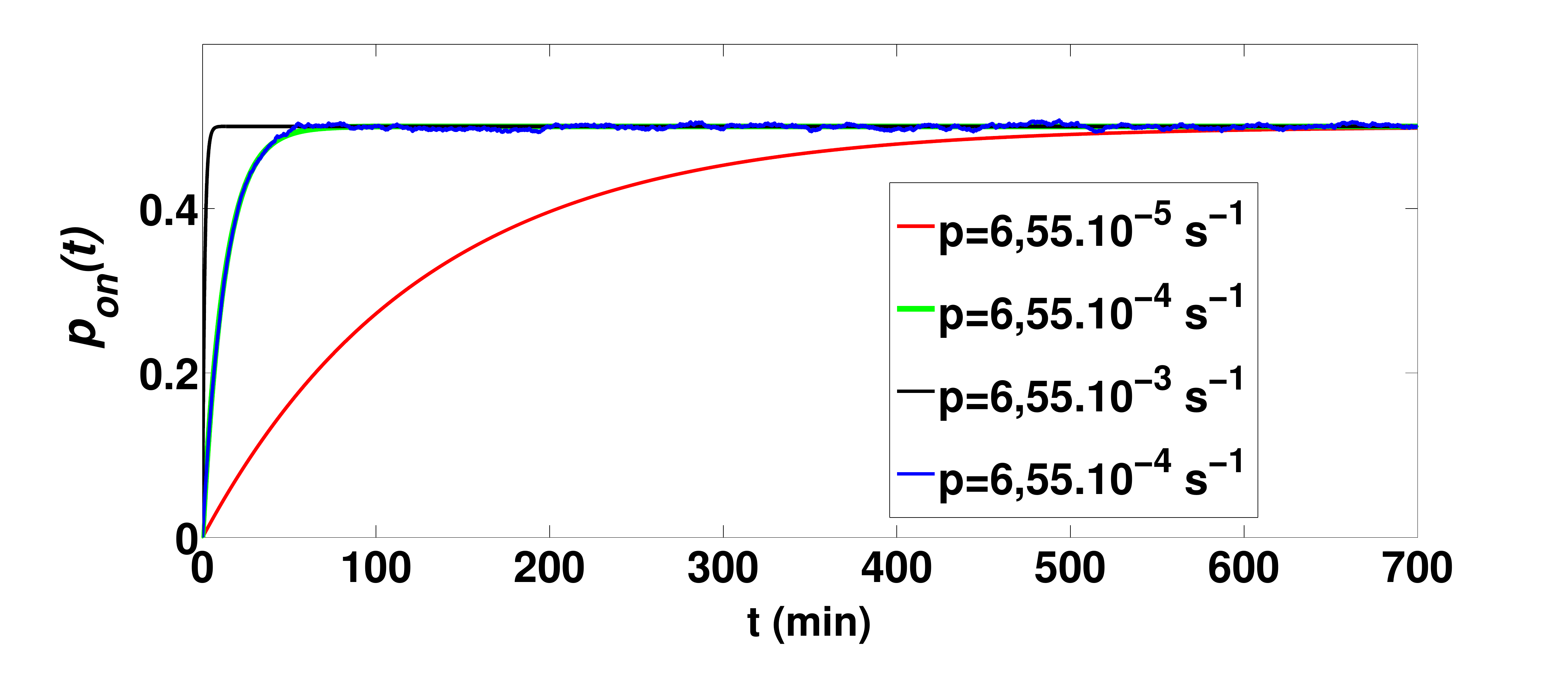}
\caption{Probability $p_{\rm on}(t)$ of finding a device on as obtained from 
Eq.(\ref{probability}) for different values of $p$. The corresponding 
characteristic times to reach the stationary states are 
$\tau=125$ min (red line), $12.5$ min (green line), and $\tau=1.25$ min (black 
line). Here $p_{\rm on}(0)=0$.}
\label{proba}
\end{figure}

This simple model corresponds exactly to a Markov process for a system composed 
of $N$ {\it particles} each one making transitions between 
two states (on, off) with rates $p$ and $q$ \cite{Toral}. If there is no 
interaction among devices this problem can be solved exactly, and for the case 
$p=q$ the average power demand is $\langle P \rangle =N P_0/2$, and, in the stationary regime, the size 
of the fluctuations is proportional to $\sqrt{N}$, with variance 
$\sigma_P=\sqrt{N}P_0/2$. As a matter of fact the variance of the 
fluctuations at all times is given by
\begin{equation}  
 \sigma^{2}_P(t)=NP^2_0 [p_{\rm on}(t)-p_{\rm on}(t)^2]
 \label{sigmat} 
\end{equation}
 where the probability $p_{\rm on}(t)$ of finding a device on is given by 
 \begin{equation}  
 p_{\rm on}(t)=\frac{1}{2}(1-e^{-2pt})+ p_{\rm on}(0)e^{-2pt}.
 \label{probability} 
 \end{equation}
Fig.~\ref{proba} shows the evolution of the probability of finding a device on 
as obtained from Eq.~(\ref{probability}) for different values of $p$ to 
illustrate the difference in the characteristic time $\tau={1 \over 2p}$ to 
reach the stationary state. In blue the probability computed averaging over a large number of noise realizations is shown for $p=6.55\times10^{-4}$ s$^{-1}$. 

\begin{figure} 
\includegraphics[width=0.5\textwidth]{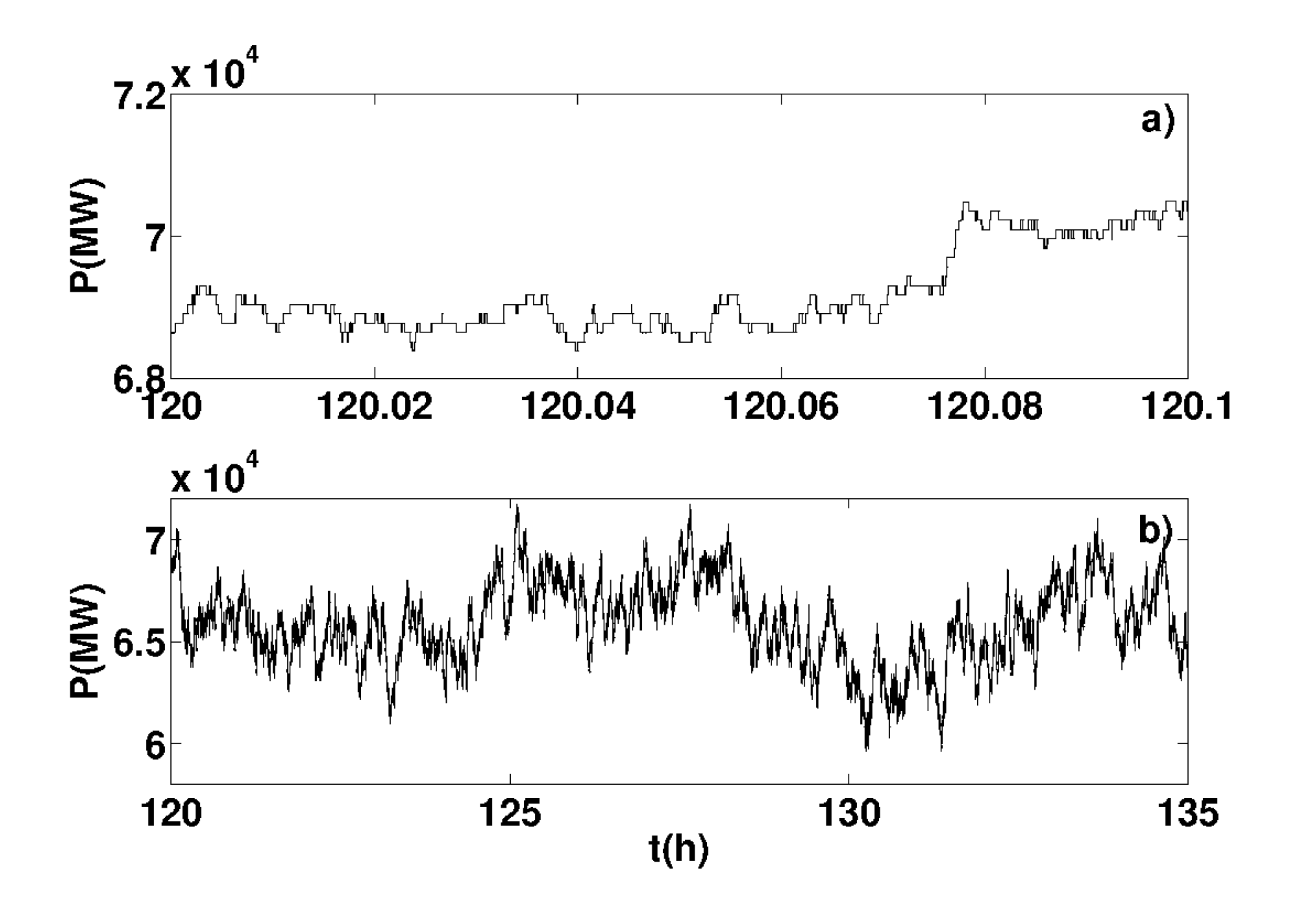}
\caption{Demand fluctuations at different time scales. Here $N=1000$, $P_0=132 MW $ 
and $p=6,55 \times 10^{-4} $ s$^{-1}$.}
\label{fig1a}
\end{figure}

We now focus on the fluctuations in the total load predicted by the model. 
Time scales play here a very important role, as the features of the 
fluctuations in Fig.~\ref{fig1a} are quite different depending on the observation time scale.
For short time scales, the above stochastic process is essentially a 
random walk, the on/off switchings of a device at each time correspond to the 
characteristic step forward or backward of a random walk. As a matter of 
fact, from (\ref{sigmat}) one can show that for short 
times the variance of the fluctuations growths as 
$\sigma_P(t) \propto \sqrt{t}$, characteristic of a random walk. 
Fig.~\ref{fig1a}a) shows a time series of the demand $P$ for $N=1000$ devices 
of power $P_0=132 MW$ with a high time resolution, where the discrete jumps can 
be clearly appreciated.     

\begin{figure} 
\centering
\includegraphics[width=0.5\textwidth]{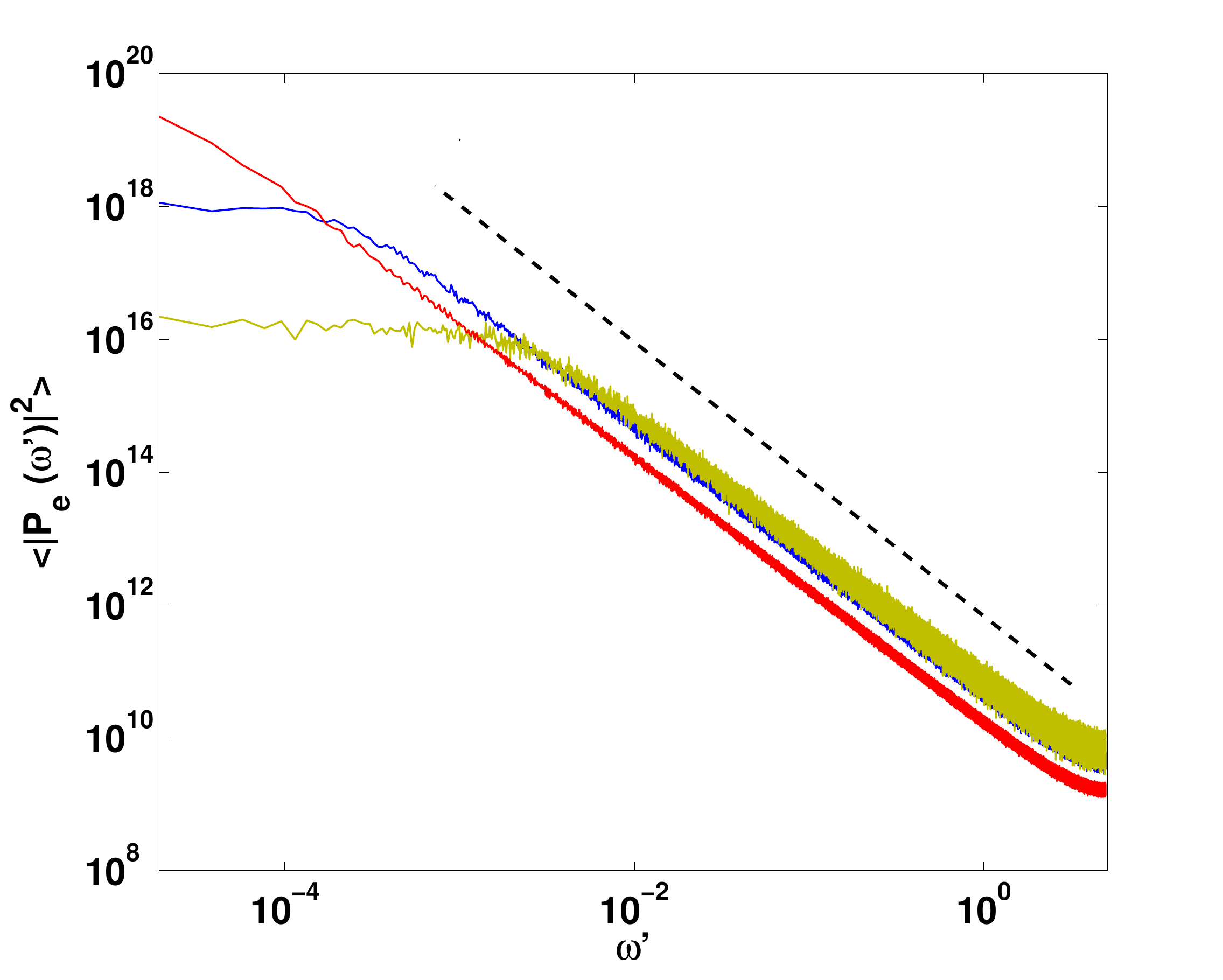}
\caption{Power spectrum of the electric load produced by the demand model for 
$p=6,55\times 10^{-5}$  s$^{-1}$ (red), $p=6,55\times10^{-4}$  s$^{-1}$ (blue), and 
$p=6,55\times10^{-3}$  s$^{-1}$ (green).
Averages over 100 noise realizations are shown. Dashed line has slope -2 for 
comparison.}
\label{powerspectrum1}
\end{figure}

However, as the number of devices is finite, the random walk is bounded, and fluctuations can not grow indefinitely. 
For long times ($t \gg \tau$) the variance saturates to 
$\sigma_P=\sqrt{N}P_0/2$. At these large time scales the fluctuations look more like 
white Gaussian noise with the latter standard deviation (Fig.~\ref{fig1a}b).
These features become clear looking at the power spectrum of the total load 
$P_{\rm e}(\omega')=\int e^{i\omega't} P_{\rm e}(t)dt$, as shown in Fig. 
\ref{powerspectrum1}.
For observations at very large time scales (low frequencies) the spectrum 
is flat, characteristic of white noise, while for short time scales (high 
frequencies) the spectrum displays a power-law shape with a slope of $-2$, characteristic of a random 
walk \cite{Toral}. 

\section{Comparison of the stochastic demand model with real measurements}
\label{results_demand}

We next compare the spectrum of the frequency fluctuations obtained from 
numerical simulations of the model with the spectrum of frequency fluctuations measured from a 
wall-plug outlet. This allows us to estimate a realistic value of the 
probability $p$.  

\begin{figure}
\centering
\includegraphics[width=0.5\textwidth]{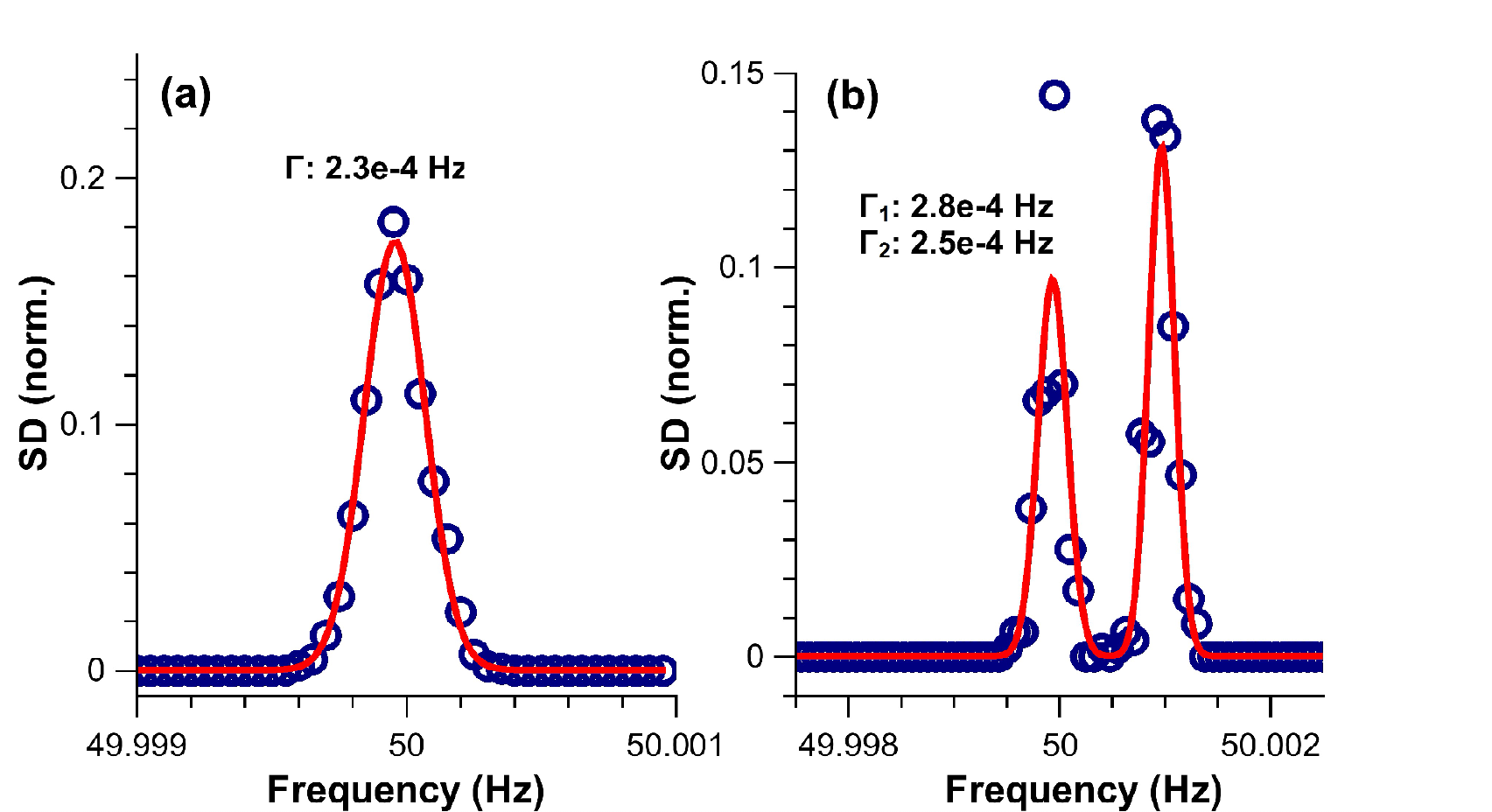}
\caption{Calibration of the spectrum analyzer. Power spectrum for a constant frequency input during $\sim$10 h (a) and for a bi-modal spectrum with a frequency separation of $1$ mHz.(b)}
\label{fig:SpectrumAnalyzer}
\end{figure}

Measurements were performed using basic custom 
electronics implemented on a Raspberry Pi 2.
 The Raspberry Pi 2 device is a small, single board and general purpose computer 
with a $900$ MHz quad-core ARM compatible CPU.
The electrical grid voltage signal was obtained from a standard wall-plug and 
was subsequently scaled to the voltage range of an analog digital converter IC 
(ADC, Texas Instruments ADC7816). The analog digital converter was
connected to the Raspberry Pi via a serial bus link. The grid signal was 
sampled using one of the CPU´s cores with a resolution of $12$ bits at a rate of 
$62.5$ KSamples/s.
The power grid waveform was recorded in real-time during temporal window of $10$ s 
which corresponds to $\approx 200$ periods of the grid signal.
 Following completion of the $10$ s sampling window, the sampling core passed the 
data to the CPU´s second core for data analysis.
As such, the device was capable to continuously sample and process the input 
waveform without loss of data.
 By measuring the temporal positions of the grid signal's zero crossings, the 
second CPU core obtained an average frequency for each $10$ s window.
 
Before its utilization as a grid frequency spectrum analyzer, we extensively 
calibrated the stability and resolution of our device, using a Keysight 33120A 
signal generator as calibration source.
 Figure \ref{fig:SpectrumAnalyzer} shows two example spectra obtained during 
calibration.
 Data shown in panel (a) corresponds to the spectral distribution during a 
$\sim 10$ hour calibration test using a $50$ Hz signal.
The obtained stability was excellent, showing no indications of drifts while 
reaching a spectral resolution of $\sim2.5 \times 10^{-4}$ Hz.
In a second calibration, we use a signal which rapidly switches from a 50 Hz to 
a 50.001 Hz.
Panel (b) shows the obtained spectra, demonstrating that the device is capable 
to clearly detect frequency changes below 1 mHz. 

\begin{figure} 
\includegraphics[width=0.5\textwidth]{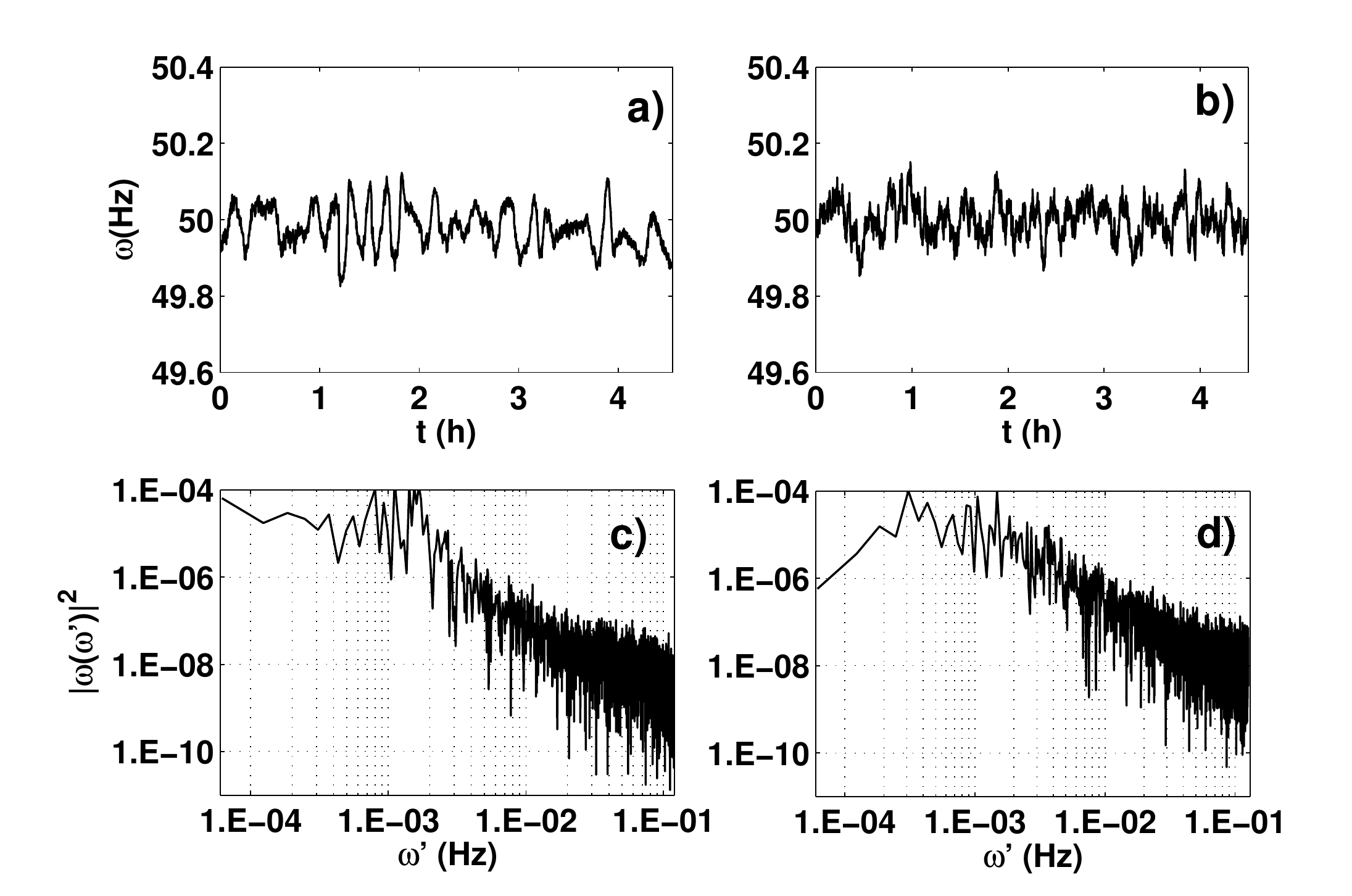}
\caption{Time series (a) and power spectrum (c) of the frequency measured at an electrical 
outlet at IFISC. Panels (b,d) show the equivalent results from a numerical simulation with $p=6.55 \times 10^{-4}$ s$^{-1}$, $N=1000$, $P_0=132 MW $ and power 
plant parameters as in Fig. \ref{fig1}.}
\label{comparison_experiment_numerics}
\end{figure}

Fig.~\ref{comparison_experiment_numerics}a) shows the experimental results for the frequency obtained measuring the output of a power outlet. One observes random fluctuations of the frequency within a range of $\pm 0.2$ Hz around the $50$ Hz reference value. Fig.~\ref{comparison_experiment_numerics}c) shows the power spectrum of the frequency. At low frequencies the spectrum is basically flat while it shows a power law decay for large frequencies. 

\begin{figure} 
\includegraphics[width=0.5\textwidth]{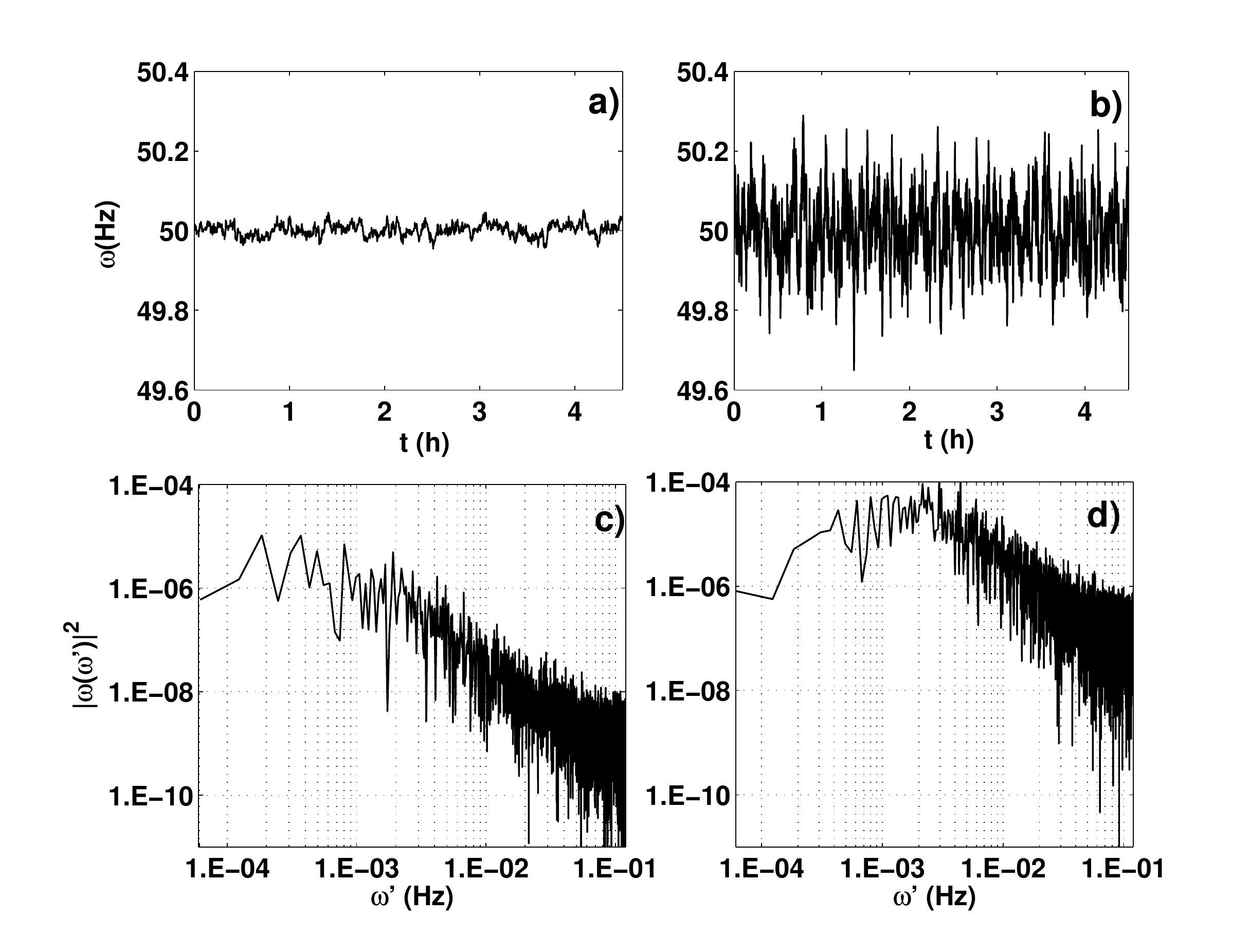}
\caption{Frequency time series and power spectrum for a numerical simulation with $p=6.55 \times 10^{-5}$ s$^{-1}$ (a,c) and  with $p=6.55 \times 10^{-3}$ s$^{-1}$ (b,d). Other parameters as in Fig. \ref{comparison_experiment_numerics}.}
\label{freqpowerspec}
\end{figure}

For comparison Fig.~\ref{comparison_experiment_numerics}b) and d) show the frequency time trace and power spectrum obtained from a simulation of Eqs.~(\ref{eq1})-(\ref{eq2}) with the stochastic demand model described in Section ~\ref{sec_demand}. Stochastic demand fluctuations as those shown in Fig.~\ref{fig1a} translate to fluctuations of the frequency (Fig. \ref{comparison_experiment_numerics}b). The power spectrum of the 
frequency fluctuations displays a plateau for low frequencies and decay as a 
power law for high frequencies (Fig. \ref{comparison_experiment_numerics}d), reproducing the characteristics of the power 
spectrum of the demand fluctuations (Fig. \ref{powerspectrum1}). Therefore, changing $p$ in the stochastic model shifts the characteristic time scales of the frequency 
fluctuations as well, which allows us to adjust the value of $p$ in order to fit the experimental data. 
Fig. \ref{freqpowerspec} shows the result generated by the model for two very different values of $p$. A shift in the frequency where the power spectrum 
starts to decay can be clearly appreciated. We finally take $p= 6.55 \times 10^{-4}$ s$^{-1}$ as a 
good value that reproduces the power spectrum of the frequency fluctuations measured experimentally (Fig.~\ref{comparison_experiment_numerics}a). 
For this probability of switching, in average, about six of the 1000 devices 
receive and order to switch on or off every 10 seconds.

We find that the adjusted probability $p$ corresponds to a characteristic time for the 
saturation of the fluctuations around 12 min. Frequency fluctuations have flat spectrum as white noise for time scales slower than this characteristic time. For faster time scales frequency variations behave as a random walk.

The experimental data displays also a superimposed periodicity, as revealed by the peak in the power 
spectrum around $1\times 10^{-3}$ Hz, corresponding to oscillations with a period of approximately
$15$ min, clearly observable in the data, and which are not introduced in the model.

\section{Dynamic Demand Control}
\label{sec_ddc}

We next include a fraction of smart devices operating under a DDC algorithm which delays the device 
switching on or off if the instant frequency is beyond a given tolerance. 
Pending tasks are recovered later in periods of favorable frequency conditions in order to ensure that 
the average energy consumption at the end of the day is the same as in absence 
of smart devices.

The proposed DDC algorithm is implemented on top of the demand model explained 
in the previous Section. It works as follows: at every time step devices can randomly turn on, off or remain in the same state they were before. However for smart devices, before committing any change of state, the DDC measures the grid
frequency $\omega$ and the change is only committed if $\omega$ is within a suitable range. Smart devices in the off state that randomly would switch on effectively do so only if the frequency is above a minimum level $\omega>\omega_{\rm R}-\epsilon$. When a switch-on is prevented by the DDC the missing consumed energy is accounted to be used at a latter time, 
frequency conditions permitting. 
Similarly smart devices in the on state that randomly would switch off effectively do so only if the frequency is below a maximum level $\omega<\omega_{\rm R}+\epsilon$, and when a switching-off is skipped, the extra consumed energy is accounted to be saved later. The objective is that, on the long run, every smart device has used the same total energy as if 
it were not smart. 

In the following we refer to the extra (saved) energy consumption generated by the DDC control as pending 
tasks.
Energy consuming pending tasks, namely, pending task that require to switching on a device to recover from a previous instance in which the device could not turn on when it was required, are recovered only if 
frequency is above a threshold $\omega>\omega_r+\epsilon_1$. Similarly pending tasks saving energy are recovered only if $\omega<\omega_r-\epsilon_1$. To avoid the simultaneous switching of all devices with pending tasks when these thresholds are 
crossed, each device starts recovering pending tasks with probability $\gamma$. 
The randomization of the response of appliances is known to avoid oscillations 
created by the synchronizations of smart devices 
\cite{Short07,Parliament,Mohsenian,Saadat}. 

Thus the overall DDC algorithm consists then of two distinct parts: decision on committing actions and recovery of 
pending tasks, and it has only three parameters, namely $\epsilon$ for the allowed range to commit actions, and  $\epsilon_1$ and $\gamma$ for the recovery of pending tasks.

\section{Effects of DDC on the power grid frequency}
\label{results_DDC}

We next analyze the role of the different DDC parameters. Typically it will be convenient to set $\epsilon$ below the statutory limit for frequency variations ($\pm 0.2$ Hz). Throughout this paper we take $\epsilon=0.05$ Hz. In this section we consider that  DDC is applied to all devices, while in section \ref {sec_fraction} we will address the effect of applying DDC only to a fraction of them.

\begin{figure} 
\includegraphics[width=0.5\textwidth]{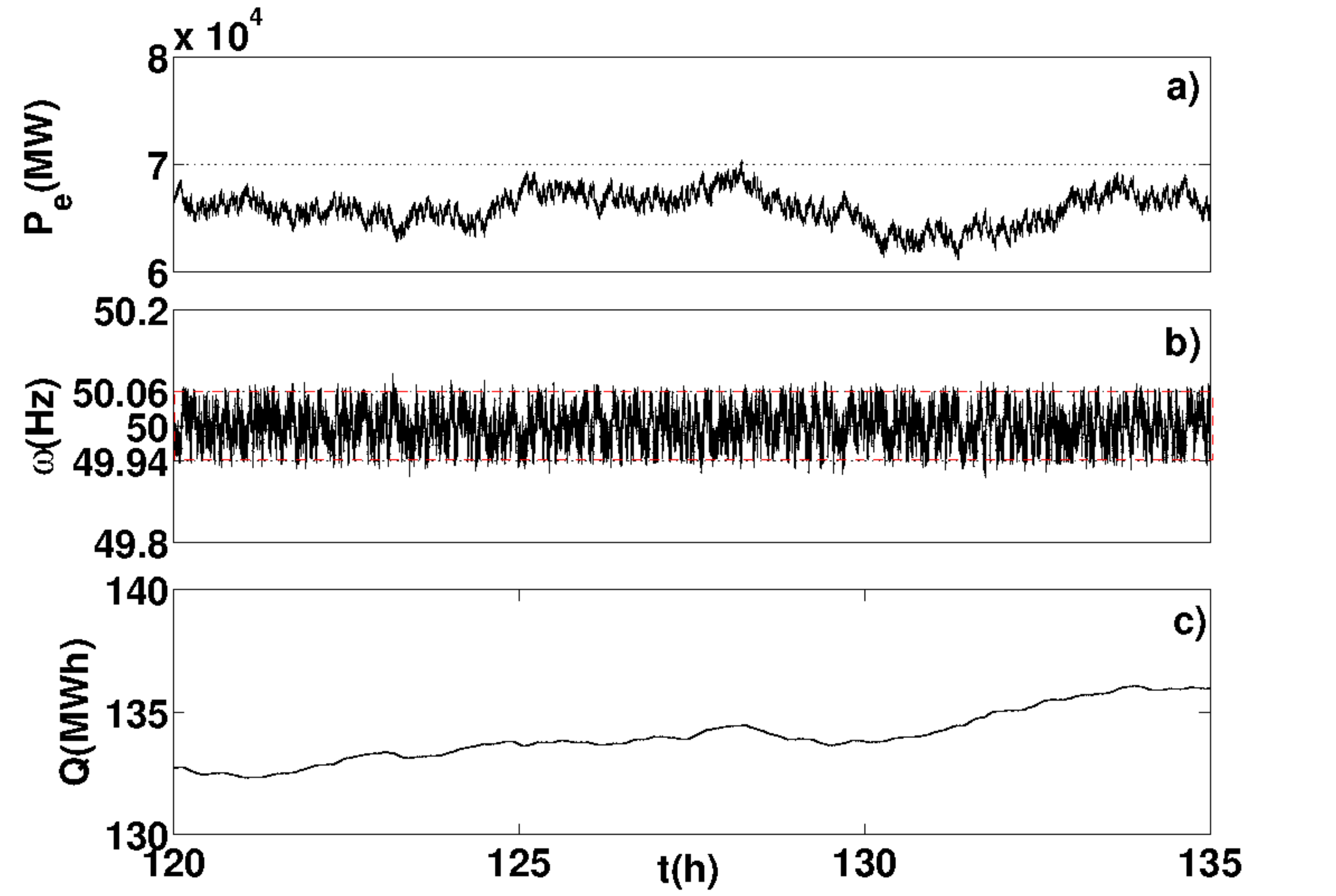}
\caption{Time series of (a) the demand $P_{\rm e}$, (b) frequency $\omega$, 
and (c) pending tasks $Q$ with DDC applied to all devices. We have considered $\epsilon=0.05$ Hz, $\epsilon_1=0.1$ Hz and $\gamma=1$. Other parameters as in Fig.~\ref{comparison_experiment_numerics} }
\label{epsilon1=0.1}
\end{figure}

We first consider a very large value of $\epsilon_1$ so that pending tasks are not recovered in practice. 
In this situation, as shown in Fig. \ref{epsilon1=0.1} the control efficiency is very good and the frequency stays almost 
always within the tolerance range $\omega_{\rm R}-\epsilon < \omega < \omega_{\rm R}+\epsilon$. 
The counterpart is, however, that depending on the system conditions smart devices will consume 
more (or less) energy than what they were supposed to consume to perform 
whatever task they were designed to do, accumulating pending tasks indefinitely, which is unrealistic. 

We define the pending tasks $Q_i$ of smart device $i$ as the absolute value of the energy that this device has 
consumed in excess or in shortage with respect to the reference case of no applying any DDC control. 
Total pending tasks on the whole grid are given by $Q=\sum_i Q_i$. 

To effectively recover pending tasks we have to decrease $\epsilon_1$. 
Decreasing the value of $\epsilon_1$ and setting recovery probability $\gamma$ to 1 leads to an effective recovery of all pending task as soon as the frequency crosses the threshold. Although this avoids the accumulation of pending tasks, it leads to  large frequency fluctuations (even larger than without DDC). All devices with energy demanding pending tasks turn on simultaneously when the frequency crosses $\omega_{\rm R} + \epsilon_1$ from below. Similarly all devices with pending tasks saving energy will simultaneously turn off when the frequency crosses $\omega_{\rm R} - \epsilon_1$ from above.

\begin{figure} 
\includegraphics[width=0.5\textwidth]{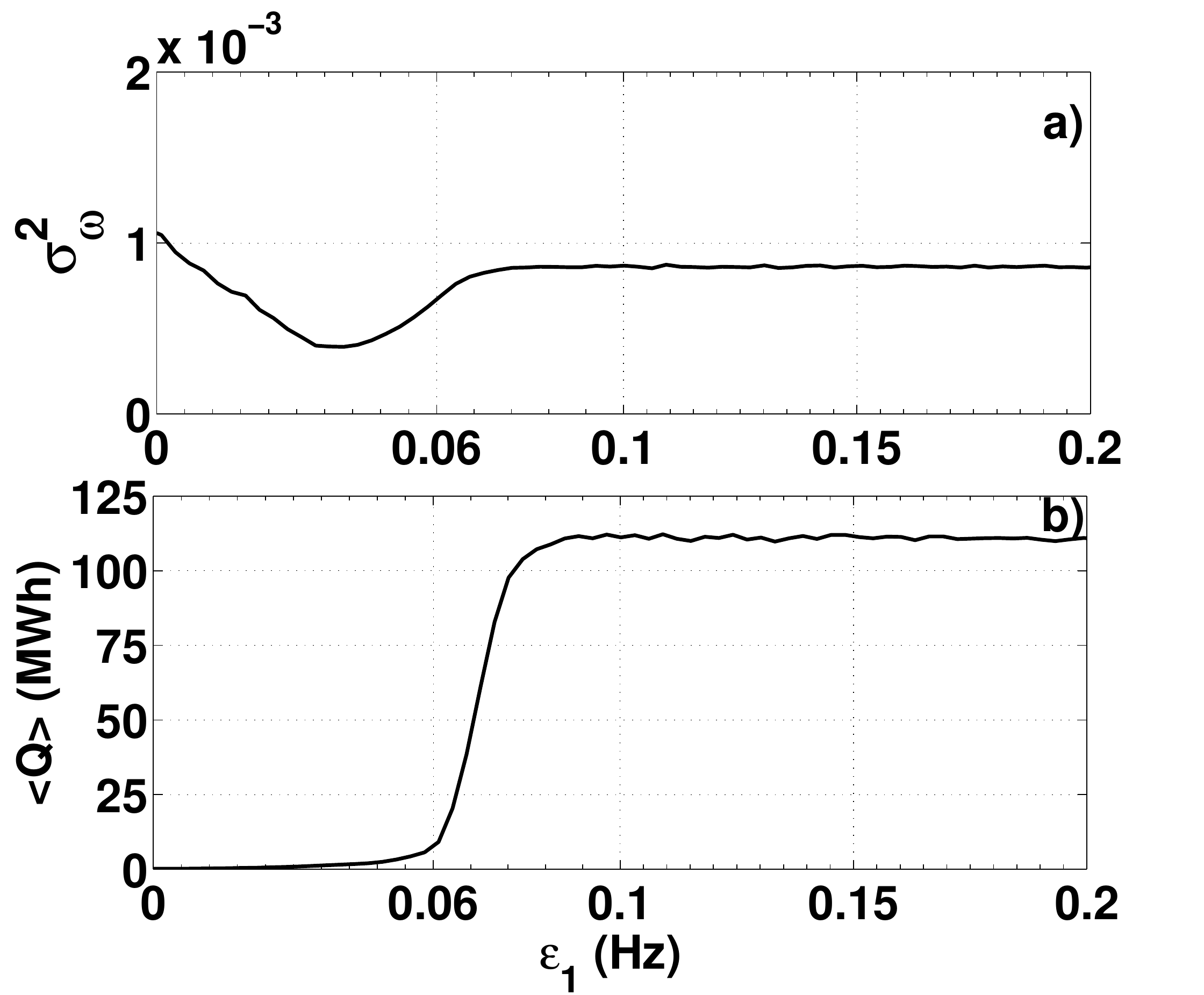}
\caption{Frequency variance (a) and pending tasks at time $t=135$ h averaged over $100$ realizations of the noise (b). 
We have considered $\gamma=1.2 \times 10^{-3}$. Other parameters as in Fig.~\ref{epsilon1=0.1}.}
\label{variance_Q_gamma=0.0012}
\end{figure}

In order to avoid this synchronized switchings it is required $0< \gamma < 1$ such that pending tasks are recovered progressively, avoiding sudden demand peaks. To determine suitable values of $\epsilon_1$ and $\gamma$ we have 
explored this two-parameter space performing numerical simulations and computing the variance of the frequency 
fluctuations $\sigma_{\omega}^2$ and number of pending tasks averaged over noise realizations $\langle Q \rangle$. 
Fig. \ref{variance_Q_gamma=0.0012} shows the results for changing $\epsilon_1$ for a fixed value of $\gamma=1.2 \times 10^{-3}$.  For 
$\epsilon_1<-\epsilon$ pending tasks are recovered practically immediately, washing out the 
effect of DDC. As $\epsilon_1$ is increased DDC starts acting, progressively reducing frequency fluctuations. 
We consider only positive values of $\epsilon_1$ as recovering pending task under unfavorable frequency conditions 
is not recommendable. For $\epsilon_1$ small the average number of pending tasks reaches a stationary value. This stationary value increases slowly with $\epsilon_1$. A qualitative change in the number of pending tasks occurs at $\epsilon_1 \approx 
0.06$ Hz, slightly above the value of the tolerance $\epsilon=0.05$ Hz of the 
frequency control. At this value pending tasks increase very sharply. 
For $\epsilon_1$ above this value the number of pending tasks does not reach an stationary value, it diverges for large times. The plateau shown in Fig. \ref{variance_Q_gamma=0.0012}b corresponds to the value of the pending tasks after a finite time, $t=135$ h, used for the numerical simulations. Physically, frequency fluctuations are quite reduced and very rarely reach this large values of $\epsilon_1$, precluding the recovering of pending tasks. In this circumstances the recovery probability $\gamma$ is practically irrelevant and the dynamics is similar to that shown in Fig.~\ref{epsilon1=0.1} for $\epsilon_1=0.1$ and $\gamma=1$. 

We note that, actually, for small values of $\gamma$ and intermediate values of $\epsilon_1$, the recovery of 
pending task helps in controlling frequency fluctuations, because additional 
devices switch on (off) only for high (low) frequencies, helping reducing 
fluctuations even further. This is the reason why the variance of the frequency 
fluctuations can be even lower that in the case without recovery of pending tasks. As a matter of fact, the variance 
of the frequency fluctuations has a minimum around $\epsilon_1=0.037$ Hz. 

\begin{figure} 
\includegraphics[width=0.5\textwidth]{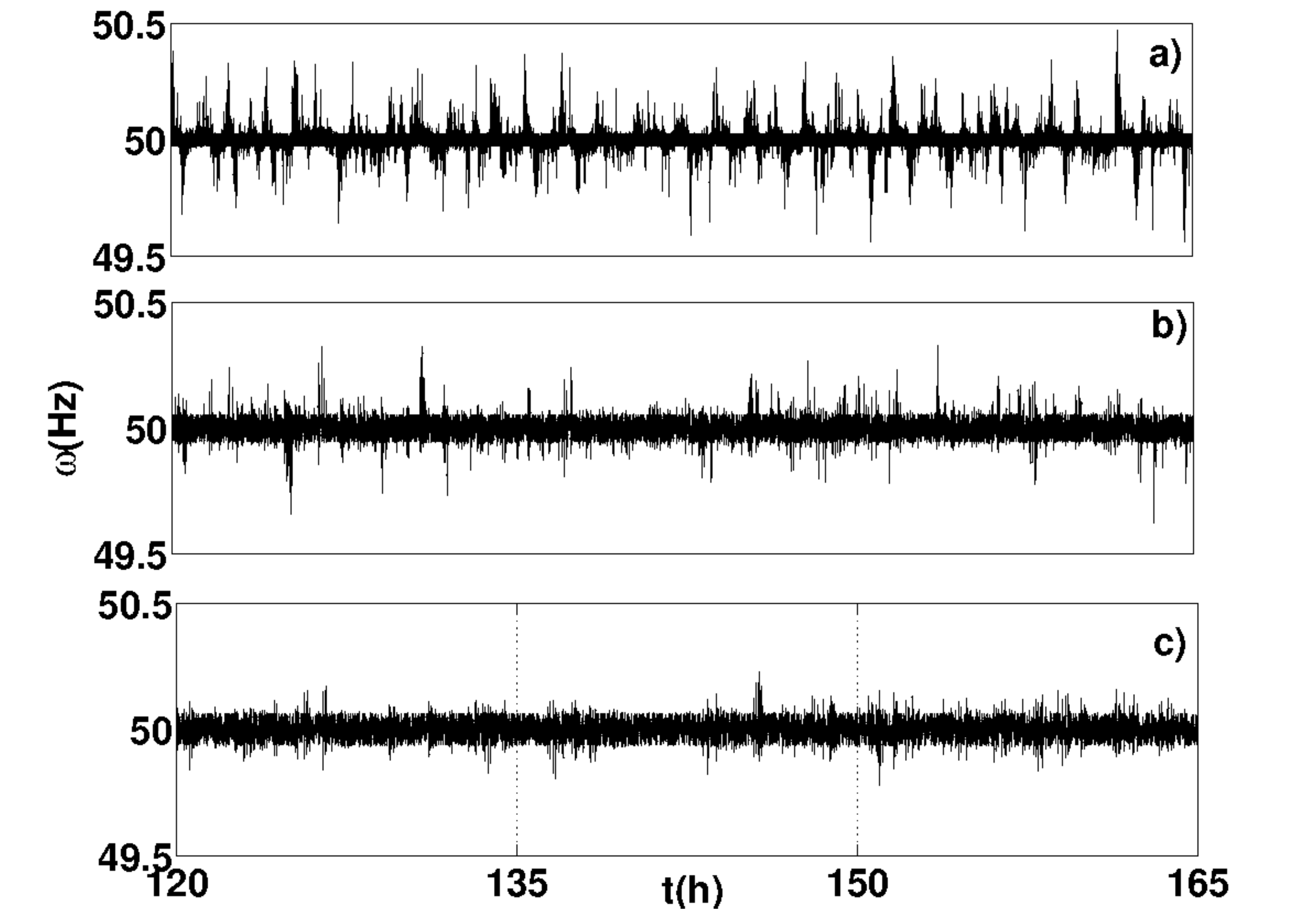}
\caption{Time series of the frequency for different values of $\epsilon_1$: a) 
$\epsilon_1=0.022$ Hz, b) $\epsilon_1=0.05$ Hz, and c) $\epsilon_1=0.06$ Hz. Other 
parameters as in Fig. \ref{epsilon1=0.1}.}
\label{trajectories}
\end{figure}

Considering the variance of the frequency fluctuations $\sigma_\omega^{2}$ and 
the number of pending tasks, from Fig.~\ref{variance_Q_gamma=0.0012} one would 
conclude that the optimum value for $\gamma=1.2 \times 10^{-3}$ is $\epsilon_1=0.037$ Hz, but this is not the whole story. Looking in 
detail at trajectories of the frequency for $\epsilon_1=0.037$ Hz, one observes 
that, despite the variance is lower than for larger values of $\epsilon_1$, 
there are extreme events in which the frequency takes very large or very small values, outside the statutory limits.
These events are rare, but pose a great risk to the system since they could trigger a 
failure or blackout. Fig. \ref{trajectories} shows trajectories of the frequency 
in simulations with different values of $\epsilon_1$. These extreme events are 
caused by the random synchronization of pending tasks recovery, and their 
probability increases decreasing $\epsilon_1$.

\begin{figure} 
\includegraphics[width=0.5\textwidth]{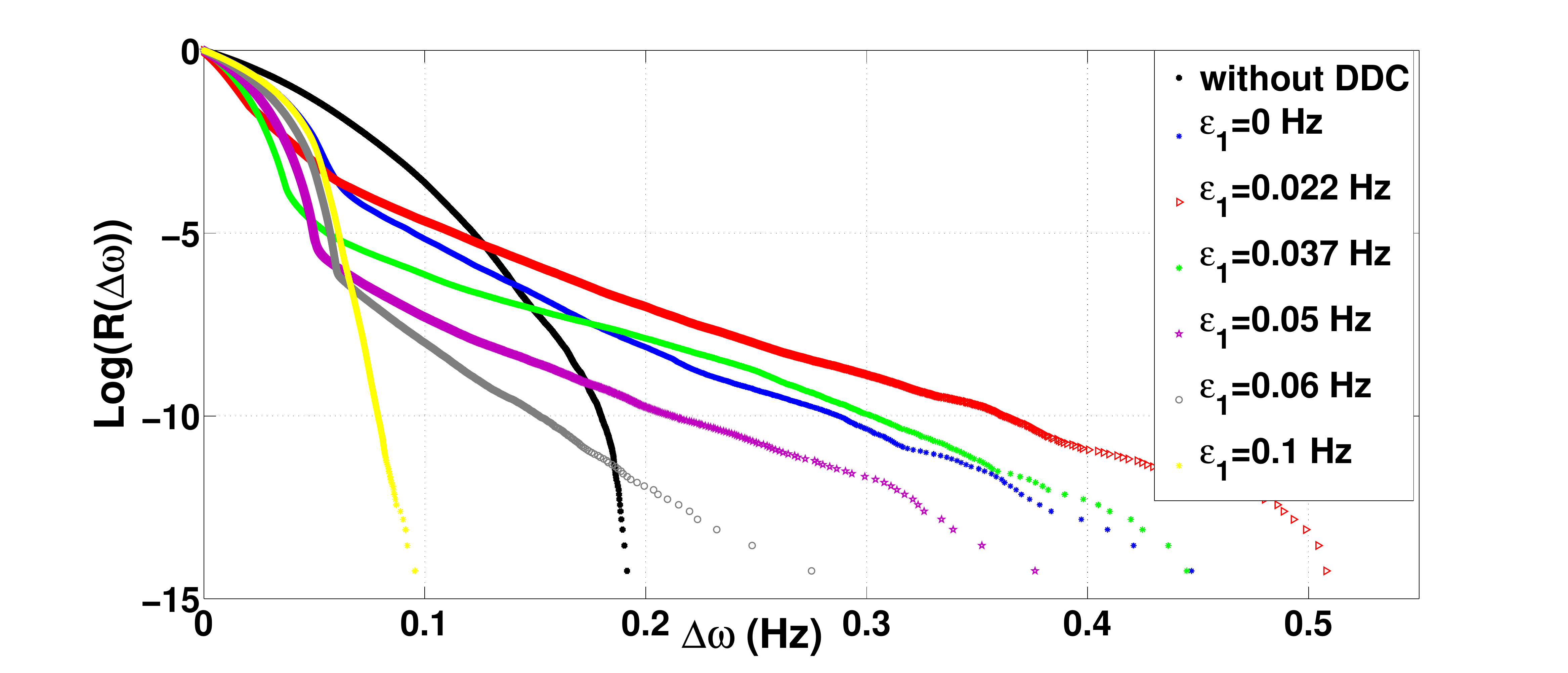}
\caption{Cumulative rank probability distribution of finding a 
fluctuation larger than $\Delta \omega= |\omega - \omega_{\rm R}|$. Parameters as in Fig. \ref{epsilon1=0.1}.}
\label{freqrank}
\end{figure}

Fig.~\ref{freqrank} shows the estimated cumulative rank probability $R$ of finding a 
fluctuation larger than $\Delta \omega= |\omega - \omega_{\rm R}|$ calculated as
\begin{equation}
 R=1-{i \over m-1},
\end{equation}
where $i$ is the rank of the frequency fluctuation of size $\Delta \omega$ in a 
very long time series with $m$ samples.
The black dots shows the case of random switching on and off without DDC for comparison. 
For the parameter considered, fluctuations above $\Delta \omega=0.2$ Hz are practically inexistent. 
When applying DDC with a large $\epsilon_1$, for instance $\epsilon_1=0.1$ Hz shown in yellow triangles in Fig. \ref{freqrank}, the frequency fluctuations are largely reduced as reflected in the narrowing of the width of the probability distribution. However, since the frequency variations rarely go above $\Delta \omega =0.1$, pending 
task are very rarely recovered and they keep accumulating. For
$\epsilon_1=0.037$ Hz, corresponding to the minimum of the variance of the 
frequency fluctuations, shown in green stars in Fig. \ref{freqrank}, the 
average number of pending tasks is low and small/medium size fluctuations are 
largely suppressed, however we observe a power-law tail of the probability 
distribution, indicating that large fluctuations have non 
negligible probabilities to occur. 

A compromise is to choose  $\epsilon_1=0.06$ Hz (gray circles in  Fig. \ref{freqrank}). For this value, we obtain a fairly 
low value for the variance of the fluctuations and average number of pending 
task, and a reasonably low probability of large events to occur, although the distribution still has power-law tails signaling that the problem of large fluctuations due to recovery of accumulated pending task does not 
disappears completely. This phenomenon is similar to what 
it is observed for the occurrence of large blackouts in power grids due to 
risk-adverse policies: strict control policies suppress small or mid-size 
blackouts but increases the probability of very large ones \cite{carreras2}. 
This is a signature of a system operating close to a critical condition 
\cite{carreras}. As a matter of fact in Fig.~\ref{freqrank} we observe a 
qualitative change in the shape of the probability distributions for 
$\Delta \omega > \epsilon_1$, signaling the recovery of accumulated pending tasks 
as the cause for the long tail of the distribution. 

\begin{figure} 
\includegraphics[width=0.5\textwidth]{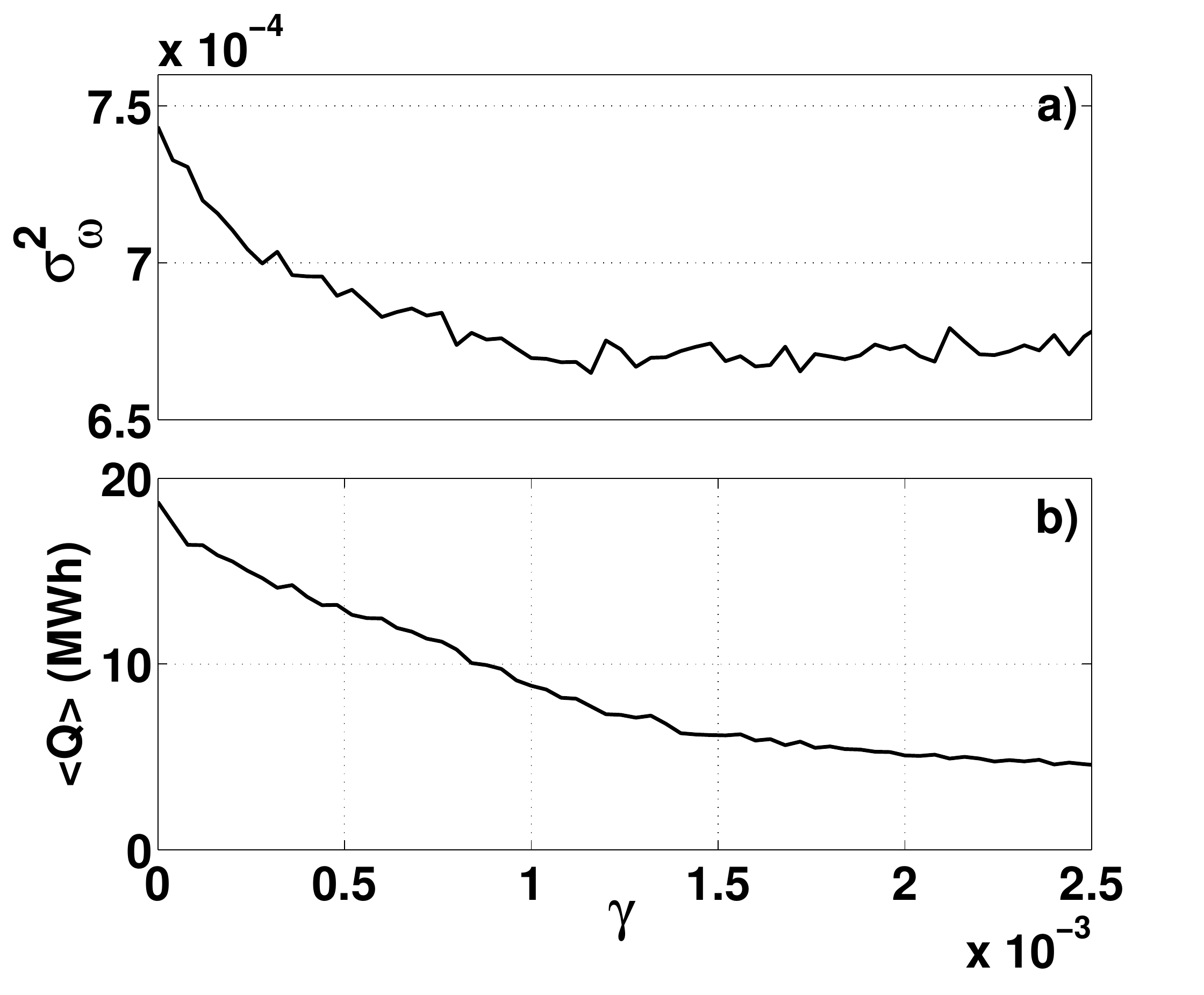}
\caption{Frequency variance (a) and average pending tasks at time $t=135$ h averaged over 100 realizations of the noise (b). We have considered
$\epsilon_1=0.06$ Hz. Other parameters as in Fig.~\ref{epsilon1=0.1}.}
\label{epsilon1=0.6}
\end{figure}

We now focus on the effect of varying the value of the recovery probability $\gamma$. 
Fig.~\ref{epsilon1=0.6} shows the variance of the frequency fluctuations and the pending tasks at time $t=135$ h 
averaged over 100 realizations of the noise as as function of  $\gamma$ for a fixed value of $\epsilon_1=0.06$ Hz. 
For low values of $\gamma$ the probability of recovering pending task is low, 
and therefore the control of the frequency fluctuations is very efficient at 
the expense of accumulating an increasing number of pending task. In fact for $\gamma = 0$, 
no task is recovered and $\langle Q \rangle$ diverges increasing the time. 
For $\gamma$ not infinitesimally small the amount of pending tasks evolves in time to a stationary value. 
As $\gamma$ is increased the stationary value for $\langle Q \rangle$
decreases rapidly without detrimental of the 
frequency variance. If $\gamma$ is too large (beyond the range of the figure) many devices start 
recovering pending task simultaneously causing synchronization peaks in the 
demand and increasing frequency fluctuations again.
We find, then, that good values for the threshold and probability of recovering 
pending tasks are $\epsilon_1=0.06$ Hz and $\gamma=1.2 \times 10^{-3}$ respectively. 
These values have been determined for a constant fixed grid load.
As a matter of fact, the optimum value of $\epsilon_1$ depends on the load of the grid. The smaller the 
load, the smaller the value of $\epsilon_1$ in order to avoid the accumulation of 
pending task, otherwise, if the load is low and $\epsilon_1$ too high, frequency 
fluctuations never reach this threshold and pending tasks can not be recovered. 
The precise dependence of $\epsilon_1$ on the system load and the effect of time varying loads will be investigated 
elsewhere.

\section{Fraction of smart devices on the grid}
\label{sec_fraction}

\begin{figure}
\centering
    \includegraphics[width=0.5\textwidth] {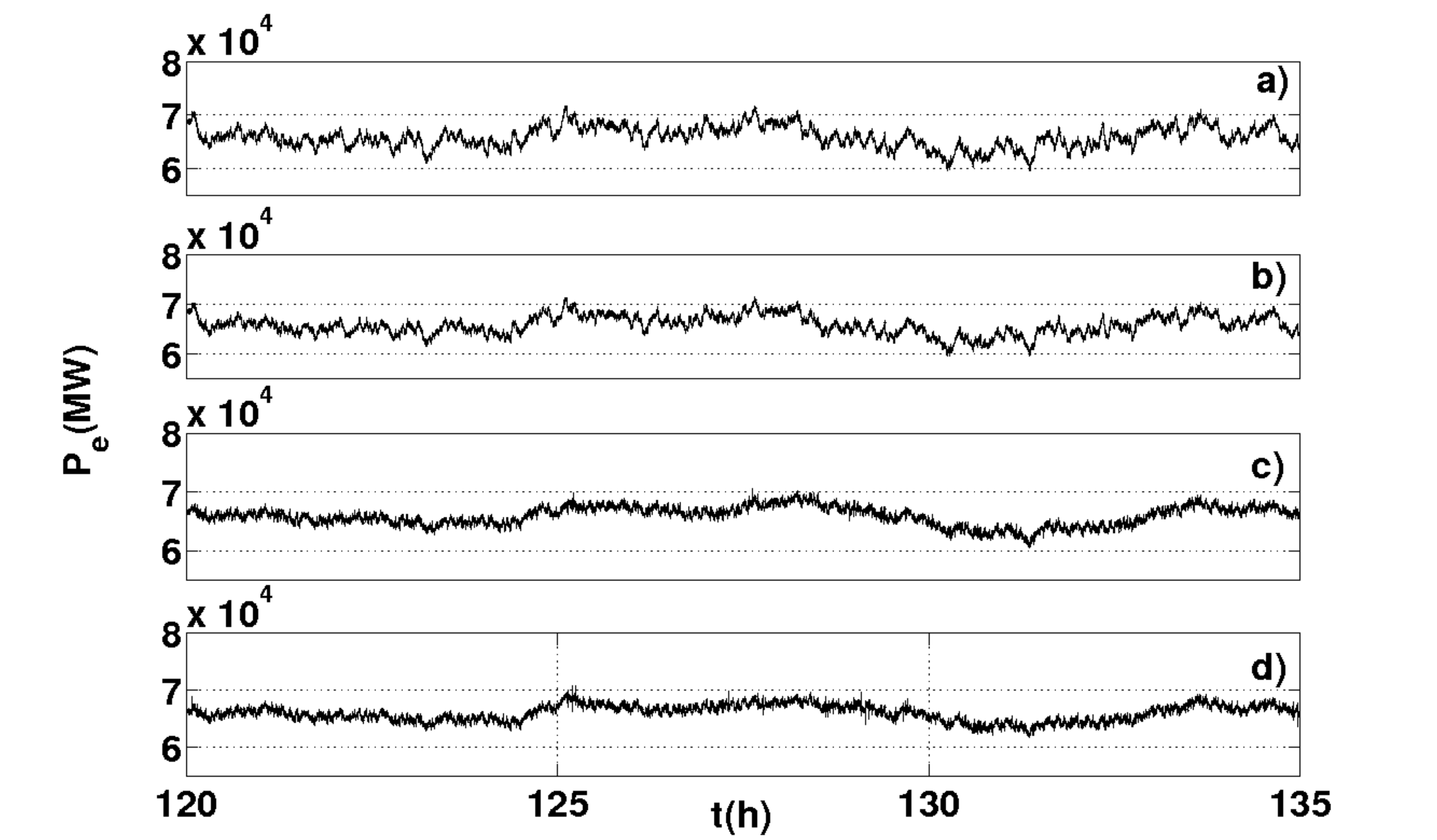} 
\caption{Time series of the demand $P_{\rm e}$ for an increasing fraction of smart devices: a) $\gamma_1=0$, b) $\gamma_1=0.01$, c) $\gamma_1=0.5$, and d) $\gamma_1=1$. 
We have consider $\gamma=1.2 \times 10^{-3}$. Other parameters as in Fig.~\ref{epsilon1=0.6}.}
\label{demandall}
\end{figure}

\begin{figure}
\centering
    \includegraphics[width=0.5\textwidth] {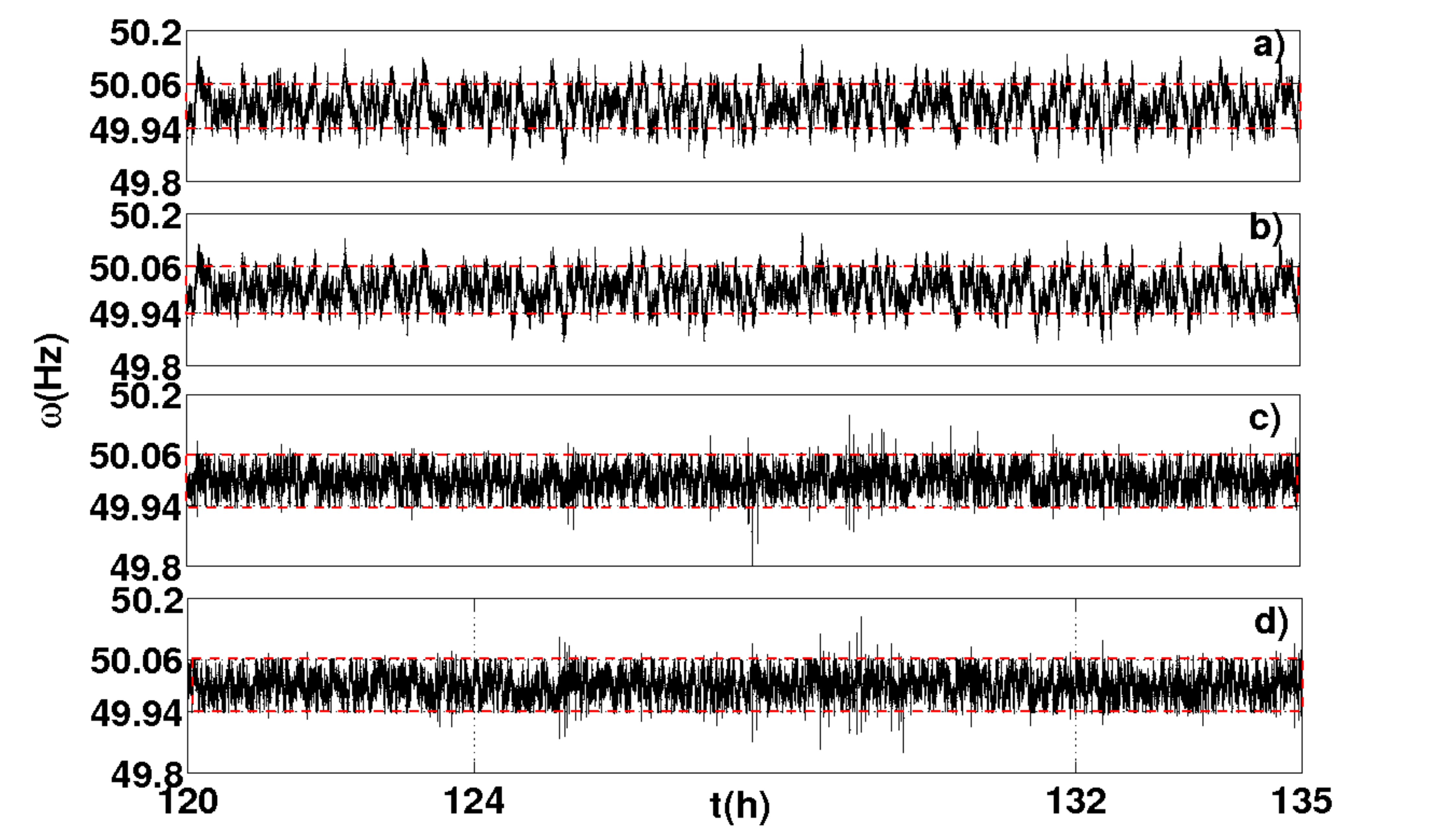} 
\caption{Time series of the frequency $\omega$ for an increasing fraction of smart devices: a) $\gamma_1=0$, b) $\gamma_1=0.01$, c) $\gamma_1=0.5$, and d) $\gamma_1=1$.
Parameters as in Fig.~\ref{demandall}.}
\label{freall}
\end{figure}

\begin{figure}
\centering
    \includegraphics[width=0.5\textwidth] {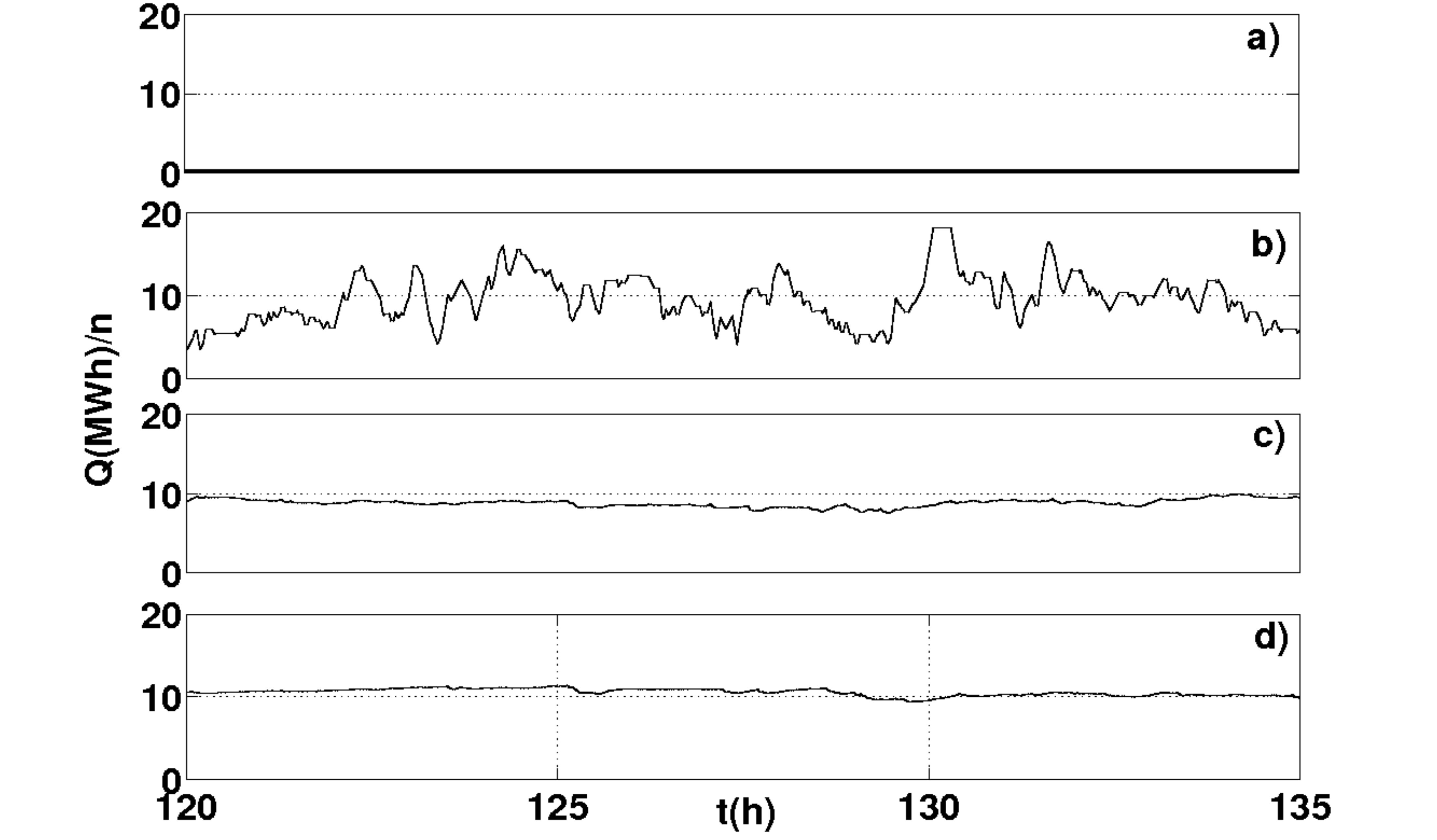} 
\caption{Time series of the pending tasks of a smart device $Q/n$ for an increasing fraction of smart devices: a) $\gamma_1=0$, b) $\gamma_1=0.01$, c) $\gamma_1=0.5$, and d) $\gamma_1=1$. Parameters as in Fig.~\ref{demandall}.}
\label{pendall}
\end{figure}

We now focus on the case in which only a fraction $\gamma_1=n/N$ of devices are smart, being $n$ the number of smart devices and $N$ the total.
So far we have considered the two extreme cases: no smart devices ($\gamma_1=0$) and all the 
load smart ($\gamma_1=1$). A more realistic case would be a grid where only part of the load 
is smart, while the rest keeps switching on and off according to random demand. In order to 
study how the performance of the grid depends on the number of smart devices, we have performed a series of simulations with the same stochastic 
realization but varying the  fraction of smart devices $\gamma_1$.
Figs. \ref{demandall}, \ref{freall}, and \ref{pendall} show time traces of the demand $P_{\rm e}$, frequency $\omega$, and pending tasks per smart device $Q/n$ for an increasing number of smart devices in the system.

\begin{figure}
\centering
    \includegraphics[width=0.5\textwidth] {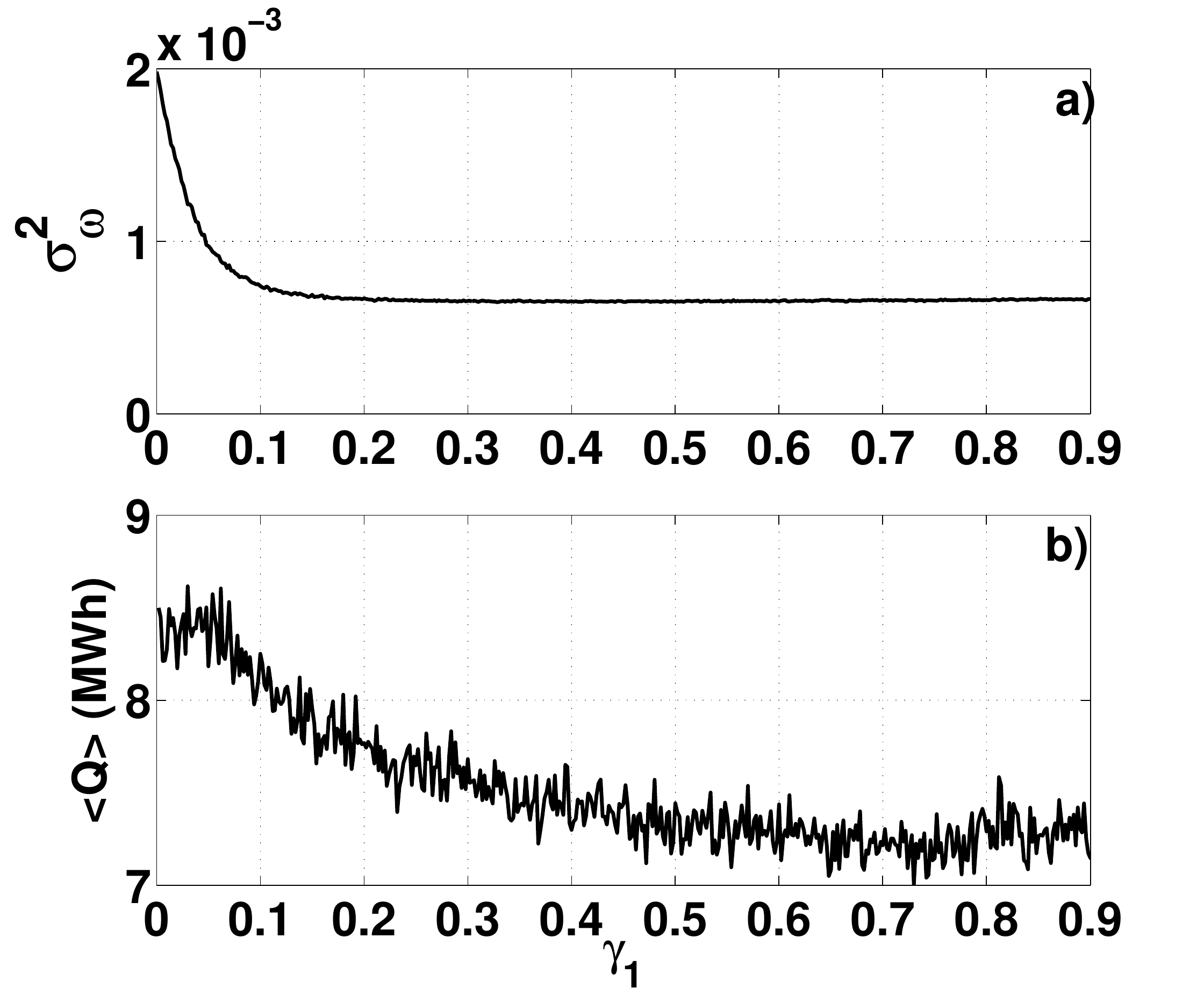} 
    \caption{Frequency  variance (a) and number of pending tasks per smart device at time $t=135$ h averaged over 100 realizations of the noise (b). Parameters as in Fig.~\ref{demandall}.}
\label{effect_gamma1}
\end{figure}

Fig.~\ref{effect_gamma1} shows the dependence of the frequency variance $\sigma^{2}_{\omega}$
and the number pending tasks per device at time $t=135$ h averaged over 100 realizations of the noise $\langle Q \rangle/n$ on the fraction of smart devices 
$\gamma_1$. We observe that the frequency variance decreases very fast increasing the fraction of smart devices. The variance saturates at $\gamma_1=0.2$. Increasing the fraction of smart devices above this value does not significantly reduces the frequency variance. 
The number of pending tasks per device decreases as well with the fraction of smart devices and saturates at a larger value $\gamma_1 \approx 0.5$. 
Thus for $0.2<\gamma_1<0.5$ while increasing the number of smart devices has a little effect on the overall frequency fluctuations it does reduce the average pending tasks. 

Since the deployment of smart devices suppose and additional cost and 
extra complexity for the appliances, and the global performance of the 
grid does not significantly increase for $\gamma_1$ above $0.2-0.4$, we conclude 
that aiming for a 30\% of the total load being smart would be a reasonable 
objective in terms of cost-benefits for the society.

\section{Conclusions}
\label{conclusions}

We have proposed a simple model to study the effects of dynamic demand control 
on the frequency of the power grid. Our model makes use of the well established equations for a power 
plant with primary and secondary regulation and introduces a simple stochastic model for the 
power demand. We have shown that the model can reproduce the statistical properties of real measurements of the frequency 
fluctuations adjusting a single parameter, namely the switching probability of the devices.

The model also allows for the application of a generic DDC protocol to a fraction of devices in order to study its effect on the dynamics. 
The generic DDC protocol consist of two parts: control, by which on or off orders on devices are actually committed only if the frequency is within a suitable range, and recovery of pending tasks only when the frequency value is appropriate and performed randomly to avoid instabilities generated by simultaneous switching of all devices with accumulated pending tasks. 

We have found that DDC can significantly reduce the variance of the fluctuations by delaying the switching 
of smart devices and recovering the pending tasks later. However, the recovery 
of pending tasks modifies qualitatively the probability distribution of the 
frequency fluctuations, introducing large tails with a power-law shape. Therefore, depending on parameters, while DDC can reduce small or 
medium size fluctuations, it can also increase the probability of observing large frequency 
fluctuations with respect to the case without control. This rare events can 
potentially trigger a failure of the system and strategies to avoid them have 
to be addressed. We have identified the most suitable parameter range for practical operation.
Finally we have also found that there is no need to apply DDC to all devices in order to achieve significant effects. Frequency fluctuations can be effectively reduced already with 20\% of devices being smart and a ratio of 30\%-- 40\% allows to reduce the fluctuations while keeping the pending tasks per device low.

\acknowledgments

We acknowledge helpful discussions with B.A. Carreras and financial support from Agencia
Estatal de Investigaci\'on (AEI, Spain) and Fondo Europeo de Desarrollo Regional under Project ESoTECoS grants number FIS2015-63628-C2-1-R (AEI/FEDER,UE), 
FIS2015-63628-C2-2-R (AEI/FEDER,UE). E.B.T.-T. also acknowledges the fellowship FIS2015-63628-CZ-Z-R under the FPI program of MINEICO, Spain.


\begin{thebibliography}{99}
\bibitem{footnote1} The role of diversity in the power of each group will be considered elsewhere.
\end{thebibliography}
\end{document}